\documentclass[letter,twocolumn]{jpsj3}
\usepackage[dvipdfmx]{graphicx}
\usepackage{amsmath}
\usepackage{amsfonts}
\usepackage{dcolumn}
\usepackage{txfonts}
\usepackage{bm}
\usepackage{physics}
\usepackage{mathtools}
\usepackage{ascmac}
\usepackage{color}
\usepackage{graphics}

\title{Unconventional \textit{s}-Wave Pairing with Point-Node--Like Gap Structure in UTe$_2$}

\author{Shingo Haruna$^1$, Takuji Nomura$^{1,2}$, and Hirono Kaneyasu$^{1,3}$}
\inst{$^1$Department of Material Science, University of Hyogo, 3-2-1 Kouto, Ako, Hyogo 678-1297, Japan \\
$^2$Synchrotron Radiation Research Center, National Institutes for Quantum Science and Technology, 1-1-1 Kouto, Sayo, Hyogo 679-5148, Japan \\
$^3$Center for Liberal Arts and Sciences, Sanyo-Onoda City University, 1-1-1 Daigakudori, Sanyo-Onoda, Yamaguchi 756-0884, Japan}

\abst{
We explore the pairing state and gap structure of $\mathrm{UTe_2}$ using a six-orbital model which we call the $f$-$d$-$p$ model. 
Our model accurately reproduces the quasi-two-dimensional Fermi surfaces consistent with recent de Haas--van Alphen oscillation measurements and the $(0, \pm \pi, 0)$ antiferromagnetic spin fluctuations observed by neutron scattering.
We incorporate on-site Coulomb repulsion for $f$ electrons and solve the linearized Eliashberg equation within the third-order perturbation theory to investigate the superconducting symmetry in $\mathrm{UTe_2}$.
The most likely state is found to be an $s$-wave state with a highly anisotropic superconducting gap structure that exhibits a point-node--like behavior of the specific heat at low temperatures.
}

\begin{document}
\maketitle

Since the discovery of superconductivity in $\mathrm{UTe_2}$ with a critical temperature $T_c = 1.6$ K \cite{Ran1,Aoki2}, this compound has been considered as a promising candidate for spin-triplet superconductors, mainly based on the following observations:
The upper critical field $H_{c2}$ appears to exceed the Pauli-limiting field \cite{Ran1,Aoki2,Knebel1,Knebel2,Sakai1}, and early nuclear magnetic resonance (NMR) measurements reported that the Knight shift along the $a$-axis does not decrease across the critical temperature $T_c$ \cite{Fujibayashi1}.
The uniform magnetic susceptibility along the $a$-axis exhibits a Curie--Weiss-like behavior, which suggests that $\mathrm{UTe_2}$ is near ferromagnetic instability \cite{Ran1}.
Owing to this magnetic susceptibility behavior,  $\mathrm{UTe_2}$ is considered as a paramagnetic end member of U-based ferromagnetic superconductors \cite{Aoki3}.
In addition, multiple superconducting phases have been discerned under magnetic fields and pressure \cite{Braithwaite1,Aoki5,Knebel3,Thomas1,Lin1,Ran4,Aoki6,Rousel1,Sakai1}.
However, recent NMR Knight shift measurements using high-quality samples ($T_c = 2.1$ K) reveal a clear decrease in the NMR Knight shift across all the crystalline axes \cite{Matsumura1}. 
Moreover, neutron scattering experiments have observed anti-ferromagnetic fluctuations at $\bm{Q} \approx (0, \pm \pi, 0)$ \cite{Butch1,Duan1}.
Furthermore, under high pressure, where a magnetic order phase is realized, the anti-ferromagnetic signals near $\bm{Q} \approx (0, \pm \pi, 0)$ are observed by the neutron scattering measurements \cite{Knafo1}.
These recent observations provide a new perspective that is markedly distinct from that in the early stage.

In terms of the band structure of $\mathrm{UTe_2}$, density functional theory (DFT)  calculations with Coulomb interaction $U$ (DFT+U, DFT+DMFT) accurately reproduce the metallic state, although DFT calculations without $U$ predict an insulating state with a band gap at the Fermi level \cite{Harima1,Ishizuka2,Xu1}.
DFT+U and DFT+DMFT calculations reproduce quasi-two-dimensional Fermi surfaces, consistent with angle-resolved photoemission spectroscopy (ARPES) experiments \cite{Miao1,Fujimori1} and de Haas--van Alphen (dHvA) effect \cite{Aoki1} measurements.
DFT+U calculations for $U = 1.5$ eV predict a Fermi pocket near the X point \cite{Ishizuka1}, which is consistent with the intensities reported by ARPES \cite{Miao1,Fujimori1}.
However, this Fermi pocket is absent in DFT+U calculations for $U=2.0$ eV, consistent with the dHvA measurements \cite{Aoki1}.

Previous theoretical studies indicate that the ferromagnetic fluctuations or the anti-ferromagnetic ones at $\bm{Q}$ lead to the spin-triplet pairing states \cite{Ishizuka1,Hakuno1,Tei1}.
In an early theoretical approach to superconductivity, possible pairing symmetry was analyzed by solving the linearized Eliashberg equation within the random phase approximation (RPA) for a six-orbital periodic Anderson model (PAM) \cite{Ishizuka1}.
The most possible state at ambient pressure was determined to be the spin-triplet $\mathrm{A_u}$ or $\mathrm{B_{3u}}$ state mediated by ferromagnetic fluctuations.
More recently, a mixed-dimensional PAM has been proposed as a minimal model for $\mathrm{UTe_2}$ that reproduces the observed Fermi surfaces and accounts for both the anti-ferromagnetic and ferromagnetic fluctuations \cite{Hakuno1}.
The analysis based on the linearized Eliashberg equation in this model suggests that spin-triplet states are possible pairing states even in the presence of anti-ferromagnetic fluctuations.
Moreover, another theoretical study revealed that Ising-like magnetic fluctuations lead to the spin-triplet superconductivity for a two-orbital model with a cylindrical Fermi surface \cite{Tei1}.

In contrast to the previous studies, we use the third-order perturbation theory (TOPT) to investigate the possibility of a pairing mechanism that is not associated with the magnetic fluctuations.
In TOPT, the effective pairing interaction is evaluated perturbatively to third order in the on-site Coulomb interaction. 
While the RPA mechanism adequately takes into account the contributions from the exchange of strong magnetic fluctuations, 
TOPT can include the contributions of third-order vertex correction terms absent in RPA \cite{Hotta1,Nomura1,Nomura2,Nishikawa1,Kaneyasu1,Yanase1,Kaneyasu2,Nomura4}. 
In contrast to the previous RPA calculations, we solve the linearized Eliashberg equation within TOPT and obtain the spin-singlet pairing state in line with the observed reduction in the NMR Knight shift.
The obtained orbital state is a highly anisotropic $s$-wave state which leads to point-node--like behavior of the specific heat in qualitative agreement with experimental observations.

Here, we introduce a model, which includes two uranium sites and two tellurium sites within a unit cell, with the same parametrization as the previously proposed PAM \cite{Ishizuka1}.
Although the setting of the hopping parameters is the same as those in Ref. \cite{Ishizuka1}, numerical values in our model are different.
The model proposed herein is called the “$f$-$d$-$p$ model", since, in our view, the term “Anderson model" should be reserved for cases without direct hoppings between the impurity-like localized $f$ orbitals.
The 5$f$ and 6$d$ orbitals at the U site and 5$p$ orbital at the Te2 site are considered, where multiplicity of those orbitals is neglected for simplicity as in Ref. \cite{Ishizuka1}.
The Hamiltonian of the $f$-$d$-$p$ model is represented as $\hat{H} = \hat{H}_0 + \hat{H}_I$, where $\hat{H}_0$ and $\hat{H}_I$ are the tight-binding Hamiltonian and the on-site Coulomb interaction between $f$ electrons, respectively.
The parameters of the tight-binding Hamiltonian are determined to reproduce the band structure given by GGA+U calculation using WIEN2k \cite{sBlaha1}, which includes the atomic spin-orbit coupling.
Detailed information regarding the tight-binding model is available in the supplemental material \cite{suppl}.
Figure \ref{fig1} (a) illustrates the crystal structure (Space group: $Immm$) and the definitions of each hopping in the tight-binding model, where the hoppings for U--U are taken to third-nearest neighbors and those for U--Te2 and Te2--Te2 are to nearest neighbors.

Diagonalizing $\hat{H}_0$ yields the band structure along the symmetry lines as depicted in Fig. \ref{fig1} (b). 
The line width in Fig. \ref{fig1} (b) indicates the weight of $f$ electrons.
The flat bands near the Fermi level predominantly consist of $f$ electrons, while the dispersive bands are primarily formed by $d$ or $p$ electrons.
In the band structure, one of the $f$-electron bands approaches the Fermi level as one moves from the $\mathrm{\Gamma}$ point to the X point.
This observation aligns with the intensities near the X point as reported by ARPES measurements \cite{Fujimori1}.  
As in Fig. \ref{fig1} (c), the $f$-$d$-$p$ model successfully reproduces the quasi-two-dimensional Fermi surfaces, where the absence of the Fermi pocket near the X point is consistent with dHvA effect measurements \cite{Aoki1}.
These Fermi surfaces comprise hole and electron surfaces, referred to as the $\alpha$ and $\beta$ surfaces, respectively. 

\begin{figure}[t]
\centering
	\includegraphics[width=0.95\columnwidth]{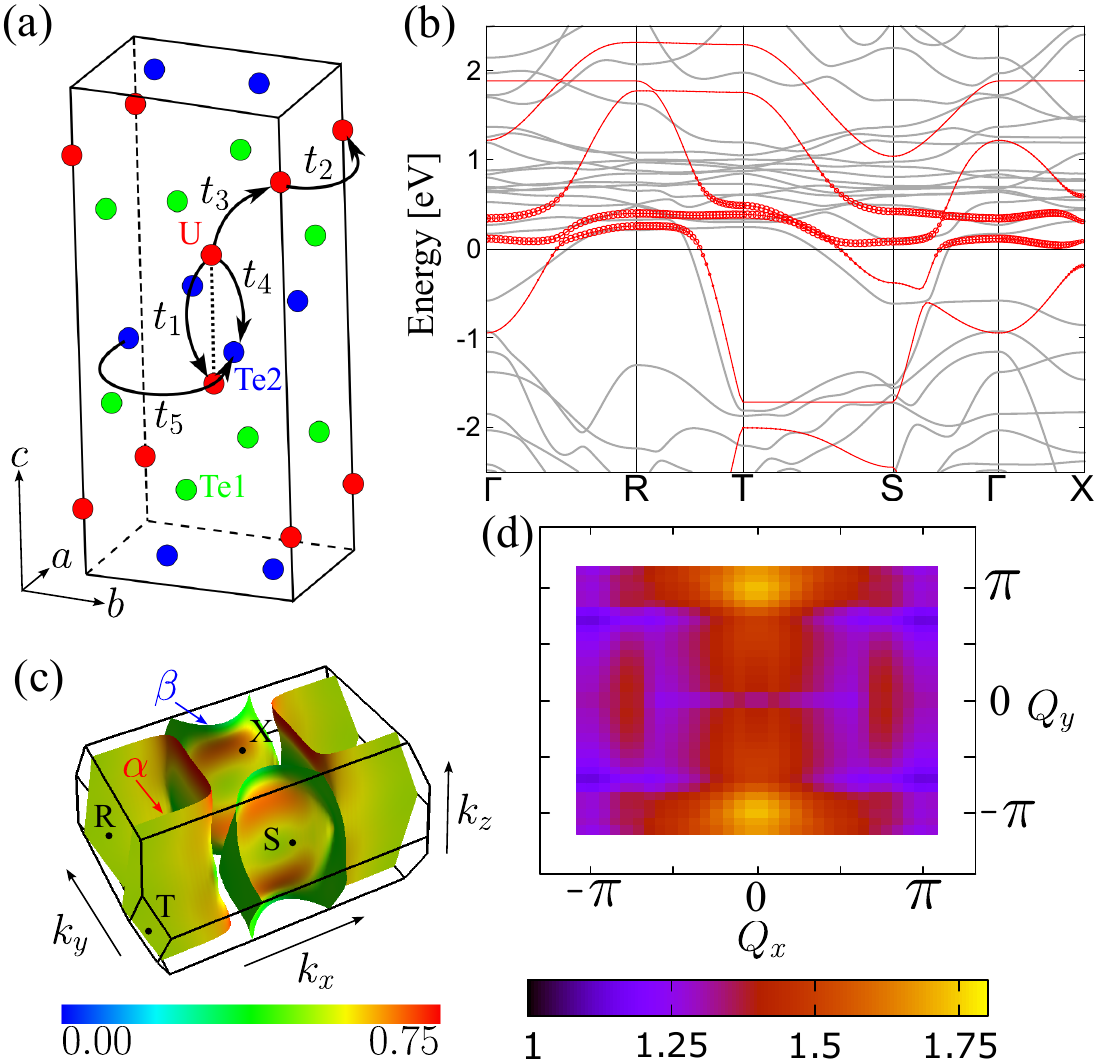}
	\caption{(a) Crystal structure (Space group: $Immm$) and $t_{1,2,3}$, $t_4$, and $t_5$ correspond to U-U, U-Te2, and Te2-Te2 hoppings, respectively.
(b) Model band structure. The line width indicates the weight of $f$ electrons.
The light grey lines display the band structure given by GGA+U ($U$ = 2.0 eV).
(c) Fermi surfaces \cite{Kawamura1}. The weight of $f$ electrons is illustrated in color. 
(d) Bare static spin susceptibility $\chi^{(0)}(\bm{Q}) = \chi^{(0)}(\bm{Q}, i\Omega_n = 0)$ on the $Q_z = 0$ plane.
$\chi^{(0)}(\bm{Q})$ has peaks at $\bm{Q} \approx (0, \pm \pi, 0)$.}
	\label{fig1}
\end{figure}

Next, we focus on the spin susceptibility, which is given as follows:
\begin{align}
\chi^{(0)}_{\mu \nu}(Q) &= \sum \limits_{\zeta_1 \zeta_2 \zeta_3 \zeta_4} \sigma^{\mu}_{\sigma_1 \sigma_4}
\chi^{(0)}_{\zeta_1 \zeta_2, \zeta_3 \zeta_4}(Q) \sigma^{\nu}_{\sigma_2 \sigma_3}, \\
\chi^{(0)}_{\zeta_1 \zeta_2, \zeta_3 \zeta_4}(Q)  &= - \frac{T}{N} \sum \limits_{k} \mathcal{G}^{(0)}_{\zeta_3 \zeta_1}(k) \mathcal{G}^{(0)}_{\zeta_4 \zeta_2}(k+Q) \label{CF},
\end{align}
where $\mathcal{G}^{(0)}(k)$ denotes the thermal Green's function for non-interacting particles, and $\sigma^{\mu}$ represents the $\mu$ component of the Pauli matrices. 
$\zeta_i \equiv (l_i, \sigma_i)$ is the combined index for orbital ($l_i$) and spin ($\sigma_i$). 
The four-dimensional momentum arguments are $k \equiv (\bm{k}, i \omega_n)$ and $Q = (\bm{Q}, i \Omega_n)$, where $\omega_n \equiv (2n +1)\pi T$ and $\Omega_n \equiv 2n \pi T$ are the fermionic and bosonic Matsubara frequencies, respectively.
Fig. \ref{fig1} (d) displays the calculated bare static spin susceptibility $\chi^{(0)}(\bm{Q}) = \chi^{(0)}(\bm{Q}, i\Omega_n = 0)$ in the $Q_z = 0$ plane.
The peaks at $\bm{Q} \approx (0, \pm \pi, 0)$ are consistent with neutron scattering experiments \cite{Butch1,Duan1}.
The $(0, \pm \pi, 0)$ anti-ferromagnetic peaks are likely due to the nesting between regions with rich $f$ components on the $\beta$ surface.

\begin{figure}[b]
\centering
\includegraphics[width=1.0\columnwidth]{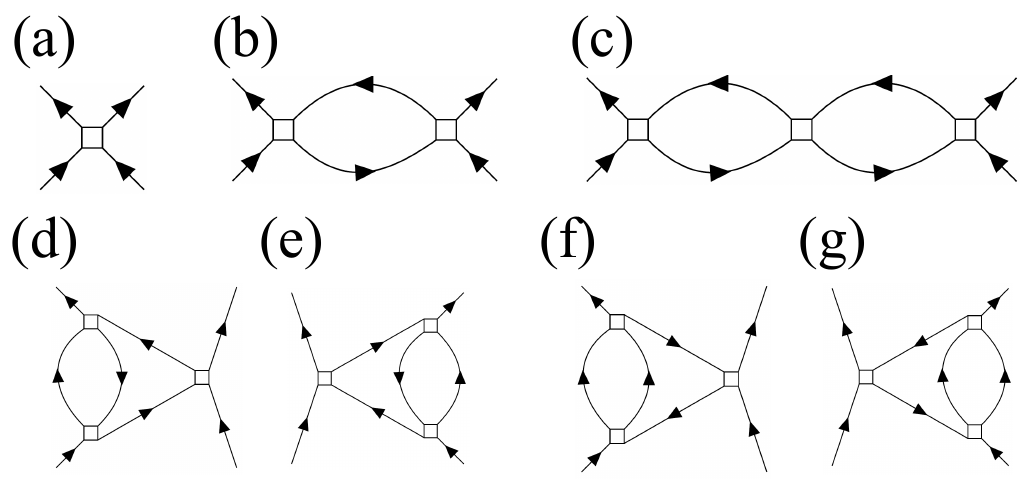}
\caption{Feynman diagrams for the effective pairing interaction in TOPT.
(a) -- (c) Diagrams are incorporated also in RPA.
(d) -- (g) Diagrams representing vertex correction terms are not included in RPA.
The empty square symbols represent the anti-symmetrized bare Coulomb interaction.}
\label{fig2}
\end{figure}

To investigate possible pairing states in $\mathrm{UTe_2}$, we solve the linearized Eliashberg equation, expressed as:
\begin{align}
\lambda \Delta_{a}(k) = - \frac{T}{N} \sum \limits_{k' a'} V^{S/T}_{a a'}(k, k') \left| \mathcal{G}^{(0)}_{a'}(k') \right|^2 \Delta_{a'}(k'),
\label{LEE}
\end{align}
where $\lambda$ and $\Delta_{a}(k)$ represent the eigenvalue and anomalous self-energy, respectively.
$\mathcal{G}^{(0)}_{a}(k)$ is the thermal Green's function for the diagonalized band $a$, and $V^{S/T}_{aa'}(k, k')$ is the effective pairing interaction, which represents the pair scattering amplitude from $(k', -k')$ to $(k, -k)$ for singlet/triplet channels.
In TOPT, we evaluate the effective interaction to third order in on-site Coulomb repulsion $U$ for $f$ electrons, as presented diagramatically in Fig. \ref{fig2}.
While the diagrams in Fig. \ref{fig2} (a) -- (c) are incorporated in the interactions considered also within the RPA, the diagrams in Fig. \ref{fig2} (d) -- (g), referred to as vertex correction terms, are not included in the RPA calculations.
The vertex correction terms can provide momentum-dependent contributions which are not mediated by magnetic fluctuations, despite TOPT being a finite-order approximation.
The momentum dependency of the eigenfunction $\Delta_{a}(k)$ determines the pairing symmetry and superconducting gap structure.
The superconducting transition point is determined when the maximum eigenvalue reaches unity.
Therefore, comparing the magnitudes of eigenvalues enables us to identify the most likely pairing state.

\begin{figure}[t]
\centering
\includegraphics[width=0.9\columnwidth]{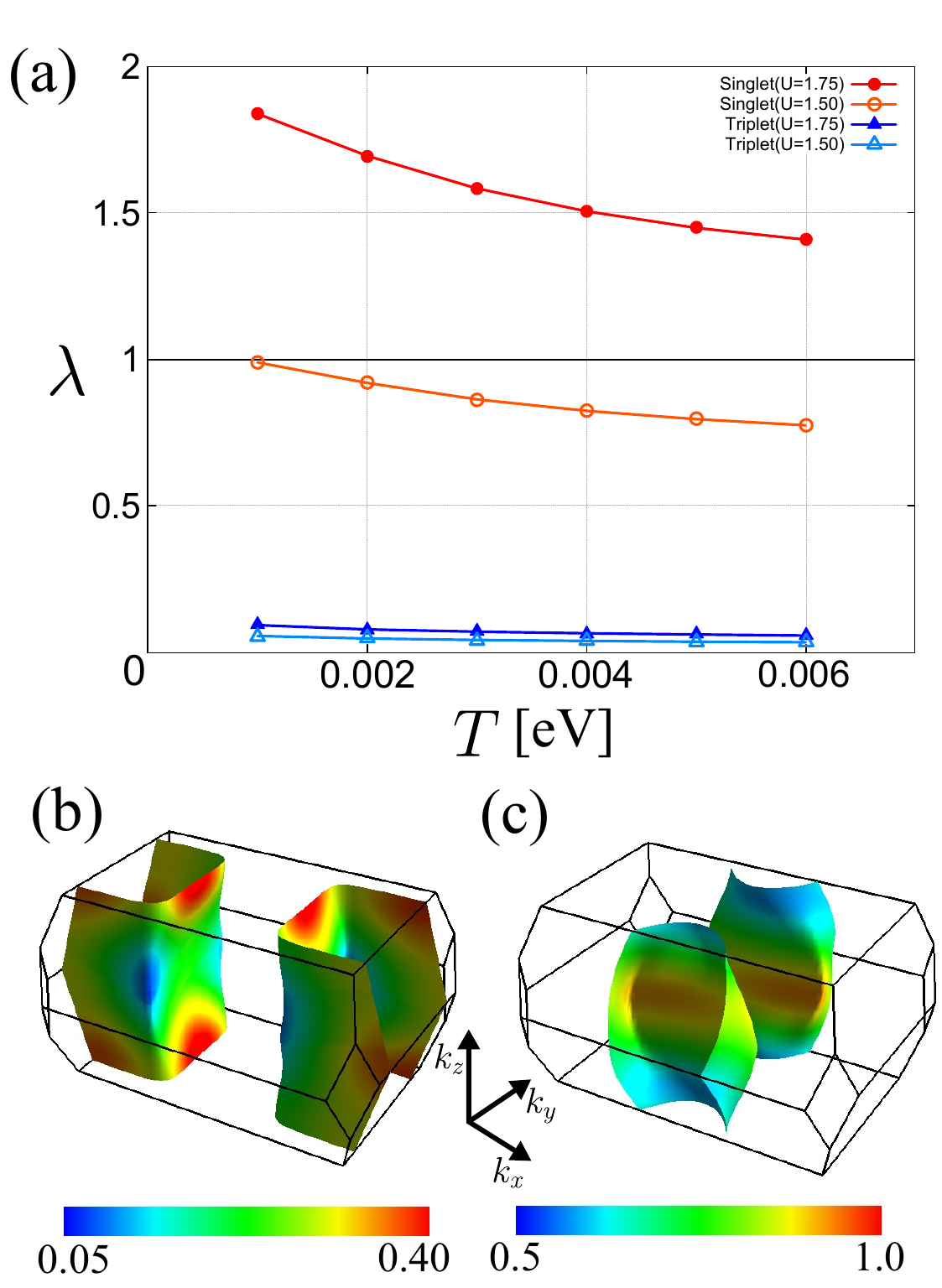}
\caption{(a) Eigenvalues of spin-singlet (circles) and triplet (triangles) states. The empty and filled symbols denote the results for $U=1.50$ and 1.75 eV, respectively.
(b, c) Magnitude of the anomalous self-energy on the $\alpha$ and $\beta$ Fermi surfaces, expressed in color.}
\label{fig3}
\end{figure}

The calculated maximum eigenvalues for spin-singlet and triplet pairings are depicted in Fig. \ref{fig3} for the Coulomb interaction $U=1.50$ and 1.75 eV as a function of temperature. 
These Coulomb interactions are smaller than the bandwidth of 2 eV, allowing for reliable perturbative treatments.
Maximum eigenvalues are given by the spin-singlet states for both the $U$ values.
Spin susceptibility in the case of spin-singlet pairing states is expected to decrease below $T_c$ along all the crystalline axes, in agreement with the reduction in the Knight shift observed in NMR measurements \cite{Matsumura1}.
To discuss the orbital state of Cooper pairs, we consider the momentum-dependency of the anomalous self-energy, $\Delta_{a}(\bm{k}, i \pi T)$, for the spin-singlet state at $T=0.001$  eV for $U=1.75$ eV, as illustrated in Fig. \ref{fig3} (b) and (c).
Although the anomalous self-energy does not change sign over these Fermi surfaces, it is highly anisotropic in the $\bm{k}$-space, exhibiting a point-node--like structure on the edges of the $\alpha$ surface at $k_z=0$.
Thus, the most likely superconducting pairing state is a highly anisotropic $s$-wave state.

\begin{figure}[b]
\centering
\includegraphics[width=0.9\columnwidth]{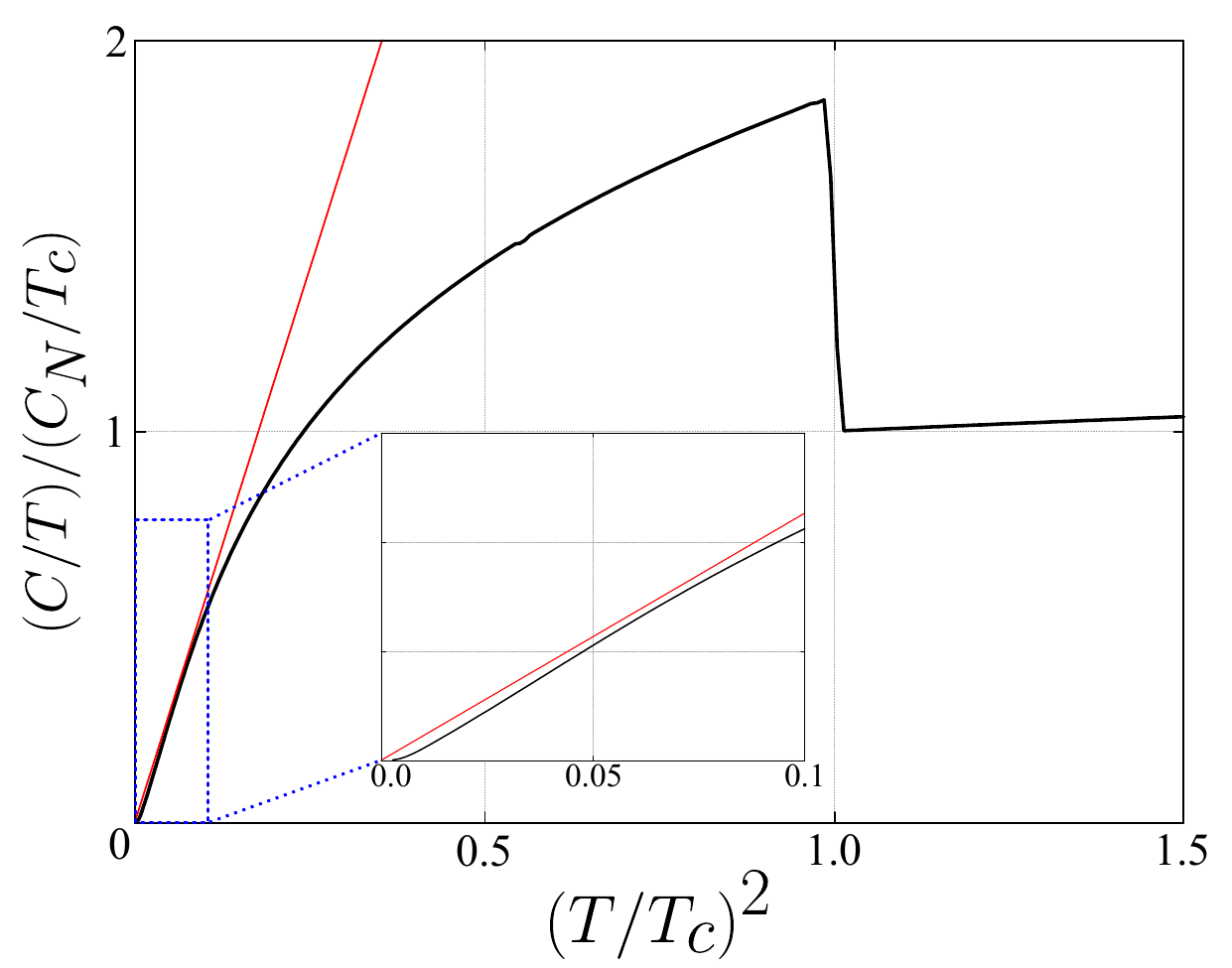}
\caption{Calculated specific heat as a function of $(T/T_c)^2$. The linear behavior is qualitatively consistent with experiments.
The vertical axis is normalized by the normal state value at $T_c$.
The $T^3$ dependence of $C$ (red line) indicates the existence of a point-node structure.
The inset shows an enlarged view of the low-temperature behavior.} 
\label{fig4}
\end{figure}

In our study, we calculate the specific heat below $T_c$ within the BCS approximation to study the relation between the point-node--like gap structure and the $T$-dependency of the specific heat. 
We assume that the gap function is expressed by the product of the momentum and temperature dependent parts, $\Delta_{a}(\bm{k}, T) = \Delta_{a}(\bm{k}) \Delta(T)$, where $\Delta_{a}(\bm{k})$ is determined above.
$\Delta(T)$ is calculated by solving the BCS gap equation \cite{Nomura3}.
The numerical results of the specific heat are depicted in Fig. \ref{fig4}.
The calculated specific heat exhibits a $T^3$ behavior in a wide range of low temperatures, which is typical of superconducting gaps with point nodes.
This temperature dependency is in qualitative agreement with experimental results \cite{Metz1,Kittaka1,Weiland1,Sakai2,Ishihara1} and is caused by low-energy excitations at the nodal region of the gap structure (see Fig \ref{fig3} (b)).
In principle, the specific heat exhibits an exponential behavior at low temperatures for the fully gapped superconducting state.
However, our calculated results demonstrate this exponential behavior of the specific heat only at extremely low temperatures, making it impossible to observe in experiments.

\begin{figure}[t]
\centering
\includegraphics[width=1.0\columnwidth]{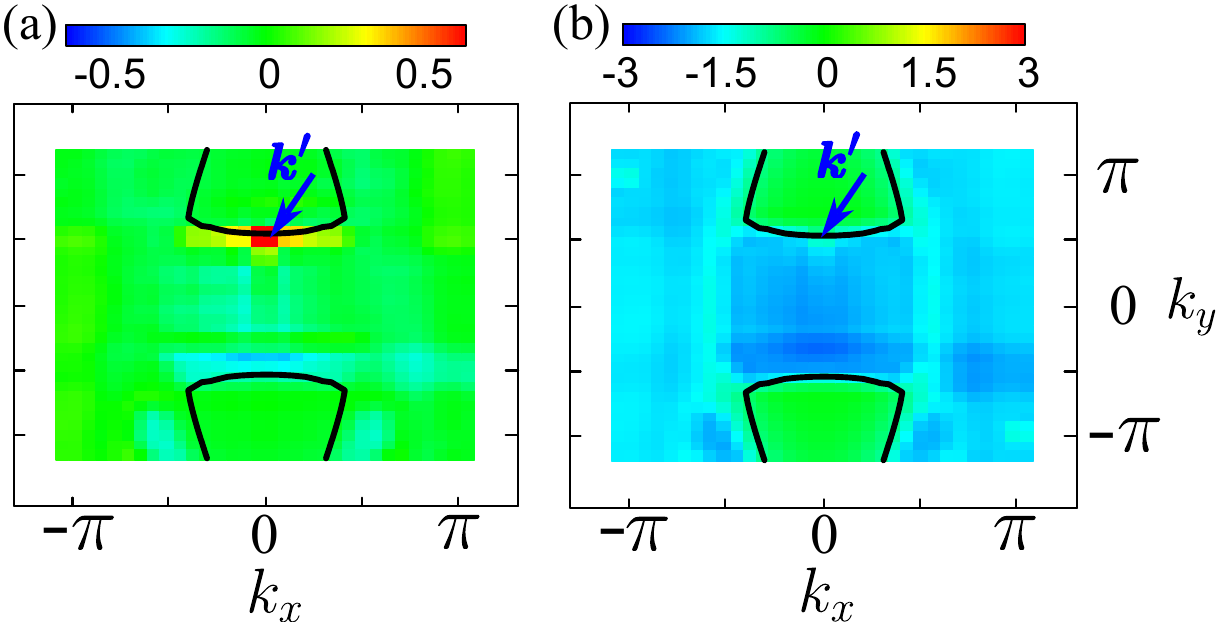}
\caption{Numerical results of the $\bm{k}$-dependency of effective pairing interaction $V^{S}_{\beta \beta}(\bm{k}~i\pi T, \bm{k}'~i\pi T)$ on the $k_z=0$ plane for the fixed $\bm{k}'$ as indicated by the arrow. Black lines indicate the $\beta$ Fermi surface. Red (Blue) indicates repulsive (attractive) region.
(a) Effective pairing interaction within TOPT.
(b) Contribution of the vertex corrections given by Fig. \ref{fig2} (f) and (g).}
\label{fig5}
\end{figure}

\begin{figure}[t]
\centering
\includegraphics[width=0.95\columnwidth]{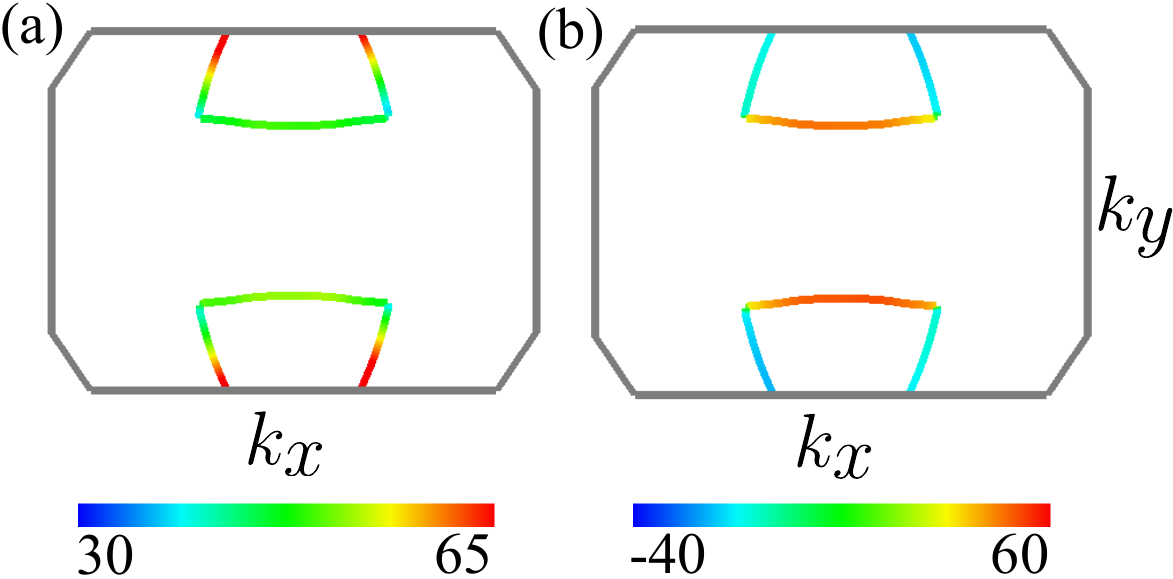}
\caption{
(a) and (b) illustrates the intra-orbital $\mathcal{F}_{ff}(\bm{k}, i \pi T)$ and inter-orbital pairing amplitude $\mathcal{F}_{ff'}(\bm{k}, i \pi T)$ on the Fermi surfaces in $k_z = 0$ plane, respectively.
The black lines indicate the Brillouin zone boundary.
Slight asymmetry in mirror reflection is seen, which is an artifact arising from the use of oblique coordinates in calculations.
}
\label{fig6}
\end{figure}

\begin{figure}[t]
\centering
\includegraphics[width=1.0\columnwidth]{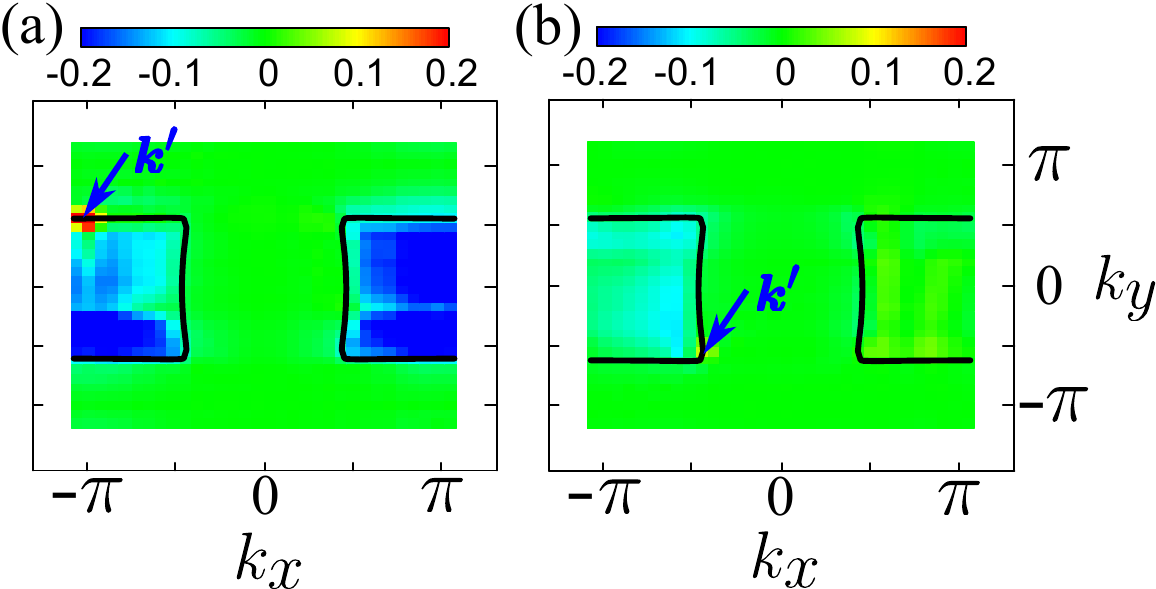}
\caption{
Momentum dependency of the effective pairing interaction $V_{\alpha \alpha}(\bm{k}~i \pi T, \bm{k}'~i \pi T)$ on the $k_z=0$ plane for the fixed $\bm{k}'$ as indicated by the arrow. Black lines indicate the $\alpha$ Fermi surface.
Red (Blue) indicates repulsive (attractive) region.
(a) The momentum $\bm{k}'$ is fixed at the largely gapped point. (b) The momentum $\bm{k}'$ is fixed at the node-like structure.
}
\label{fig7}
\end{figure}

Finally, we discuss the cause of the unconventional anisotropic $s$-wave pairing, by explicitly giving the momentum dependency of the pairing interaction.
The pairing interaction $V^{S/T}_{aa'}(\bm{k}~i \pi T, \bm{k}'~i \pi T)$ represents the pair scattering amplitude from $(\bm{k}', -\bm{k}')$ on band $a'$ to $(\bm{k}, -\bm{k})$ on band $a$ for singlet/triplet channels.
Figure \ref{fig5} illustrates the momentum dependency of the pairing interaction $V^{S}_{\beta \beta}(\bm{k}~i \pi T, \bm{k}'~i \pi T)$, where the $\bm{k}'$ is fixed on the $\beta$ Fermi surface as shown by the blue arrow.
While the effective pairing interaction is strongly repulsive around $\bm{k} \approx \bm{k}'$, it is attractive around $\bm{k} \approx - \bm{k}'$ (Fig. \ref{fig5} (a)).
This attractive interaction enables the linearized Eliashberg equation to have a non-trivial solution without sign change in the anomalous self-energy.
The attraction is caused by the vertex corrections of Fig. \ref{fig2} (f) and (g), as shown in Fig. \ref{fig5} (b), where the $\bm{k}$-dependency of the vertex corrections is extracted.
Figure \ref{fig5} (b) indicates that the vertex corrections yield the attractive interaction in a wide region of the momentum space, which is  strongly attractive particularly at $\bm{k} \approx - \bm{k}'$.
This attractive interaction gives rise to the highly anisotropic $s$-wave pairing state without sign change in the order parameter.
To investigate why the $s$-wave pairing state is favored despite of  the strong on-site Coulomb repulsion, we calculate the orbital-resolved pairing amplitude defined as $\mathcal{F}_{l_1 l_2}(\bm{k}, \tau) \equiv \ev{c_{\bm{k} l_1}(\tau) c_{-\bm{k} l_2}}$.
The calculated intra-site $\mathcal{F}_{ff}(\bm{k}, i \pi T)$ and inter-site pairing amplitude $\mathcal{F}_{ff'}(\bm{k}, i \pi T)$ on the Fermi surfaces in the $k_z=0$ plane are illustrated in Fig. \ref{fig6} (a) and (b), respectively.
As shown in Fig. \ref{fig3} (c), the superconducting gap has the maximum value on the $\beta$ surface in the $k_z = 0$ plane.
Around the $\bm{k}$ point where the gap takes the maximum value, the inter-site pairing amplitude also takes the maximum value as depicted in Fig. \ref{fig6} (b).
On the other hand, the intra-site pairing amplitude becomes small in the maximally gapped region, although it does not negligibly vanish there.
Due to this inter-site pairing mechanism, Cooper pairs can be formed in the $s$-wave channel while suppressing the on-site Coulomb energy.
Next, we discuss the origin of the point-node--like structure.
The pairing interaction on the $\alpha$-bands is illustrated in Fig. \ref{fig7}. 
In Fig. \ref{fig7} (a), the momentum $\bm{k}'$ is fixed at the largely gapped point and, the negative, i.e., attractive scattering amplitude is seen, which allows the linearized Eliashberg equation to have non-trivial solutions.
However, when the momentum $\bm{k}'$ is fixed at the nodal point, the scattering amplitude is very small on the Fermi surface as shown in Fig. \ref{fig7} (b) and it suggests that the nodal points neither contribute to the positive nor negative pair scattering amplitude.
This means that the electrons around the nodal points do not participate in pair-formation scattering.

In summary, we have discussed the most likely superconducting state, gap structure and momentum dependence of the pairing interaction using the $f$-$d$-$p$ model.
Although the parameterization of the $f$-$d$-$p$ model is the same as that of the periodic Anderson model \cite{Ishizuka1}, the numerical values of the parameters are different.
Thus, the $f$-$d$-$p$ model reproduces the quasi-two-dimensional Fermi surfaces and anti-ferromagnetic fluctuations in good agreement with dHvA \cite{Aoki1} and neutron scattering \cite{Duan1,Butch1} experiments, respectively.
To investigate the superconductivity, we have solved the linearized Eliashberg equation within TOPT.
Although our investigation was restricted to the case without magnetic fields under ambient pressure, we have obtained the highly anisotropic $s$-wave state with point-node--like gap structure, which yields the $T^3$ behavior of the specific heat at low temperatures in qualitative agreement with experiments \cite{Metz1,Kittaka1,Weiland1,Sakai2,Ishihara1}.
The attractive component of the pairing interaction, which leads to this unconventional anisotropic $s$-wave pairing, arises from the third-order vertex corrections, although the RPA-like contributions of magnetic fluctuations are included up to third order.
The spin-singlet states align with the recent NMR Knight shift measurements \cite{Matsumura1}.
Our results suggest that the possibility of spin-singlet pairing states should not be entirely dismissed yet, particularly in zero-field, and further investigation and verification of the pairing state are required.

\begin{acknowledgment}
We are grateful to J. Ishizuka, Y. Yanase and S. Fujimori for valuable and stimulating discussions.
We gratefully acknowledge the financial support from the Iketani Science and Technology Foundation.
\end{acknowledgment}

\nocite{*}
\bibliographystyle{jpsj}
\bibliography{Ref}

\clearpage
\onecolumn

\centerline{{\large \textbf{Erratum:}}}
\centerline{{\large \textbf{Unconventional \textit{s}-Wave Pairing with Point-Node--Like Gap Structure in UTe$_2$}}}

\vspace{10pt}

\centerline{{Shingo Haruna$^1$, Takuji Nomura$^{1,2}$, and Hirono Kaneyasu$^{1,3}$}}
\centerline{ \textit{
$^1$Department of Material Science, University of Hyogo, 3-2-1 Kouto, Ako, Hyogo 678-1297, Japan
}}
\centerline{ \textit{
$^2$Synchrotron Radiation Research Center, National Institutes for Quantum
}}
\centerline{ \textit{
Science and Technology, 1-1-1 Kouto, Sayo, Hyogo 679-5148, Japan
}}
\centerline{ \textit{
$^3$Center for Liberal Arts and Sciences, Sanyo-Onoda City University,
}}
\centerline{ \textit{
1-1-1 Daigakudori, Sanyo-Onoda, Yamaguchi 756-0884, Japan
}}

\vspace{10pt}

To prepare figures of some calculated $\bm{k}$-dependent quantities, such a quantity $f(\bm{k})$ on a specific set of generally oblique $\bm{k}$-mesh points has to be transformed to $f(\tilde{\bm{k}})$ on another arbitrary set of denser or Cartesian $\tilde{\bm{k}}$-mesh points.
To do this, we transform the original data $f(\bm{k})$ to the Fourier coefficients $c_{n_1 n_2 n_3}$, and then we calculate $f(\tilde{\bm{k}})$ by performing inverse Fourier transformation of $c_{n_1 n_2 n_3}$.
In other words, the data $f(\tilde{\bm{k}})$ used for plotting are obtained by interpolating $f(\bm{k})$ on the original $\bm{k}$-mesh points.
However, due to an error in a program for the inverse Fourier transformation, Figs.~3(b), 3(c), and 6 in the original article\cite{Haruna2024} are incorrect.
Therefore, Figs.~3(b), 3(c), and 6 in the original paper have to be replaced with Figs.~3(b), 3(c), and 6 in this erratum.

In Figs.~3(b), 3(c) in our original text\cite{Haruna2024}, point-node-like structures wrongly appeared at the corners of the $\alpha$-surface in $k_z = 0$ plane.
However, correct point-node-like structures appear on the $\beta$-surface near the zone boundary of $k_z=2\pi/c$, where $c =$ 13.812\AA ~denotes the lattice constant.
Although the positions of the point-node-like structures have changed, our conclusion remains unchanged: the most likely pairing state in the $f$-$d$-$p$ model is the point-node-like $s$-wave pairing state.

Figure 6 in the original text \cite{Haruna2024} must be replaced with Fig. 6 in this erratum.
Figure 6 in this erratum shows the intra-site and inter-site pairing amplitudes of $\mathcal{F}_{ff}(\bm{k}, i \pi T) + \mathcal{F}_{f'f'}(\bm{k}, i \pi T)$ and $\mathcal{F}_{ff'}(\bm{k}, i \pi T) + \mathcal{F}_{f'f}(\bm{k}, i \pi T)$, respectively.
In Fig.~6, the pairing amplitudes are illustrated on the $k_z = 0$ and $k_z = 0.44 \times 2 \pi / c$ plane.
The pairing amplitude takes its maximum value in $k_z = 0.44 \times 2 \pi / c$ plane for both the intra-site and inter-site cases.
As shown in Fig.~6, the amplitude of the inter-site pairing appears to be comparable to that of the intra-site pairing.
In addition, after the publication of our original paper, we have obtained clear results indicating that the inter-site scattering gives rise to a strong attractive interaction, whereas the intra-site scattering does not contribute to the attractive interaction \cite{HarunaUNP}.
Therefore, the main conclusion in the original text \cite{Haruna2024} is unchanged, namely, the inter-site pairing makes an important contribution to the $s$-wave pairing in the $f$-$d$-$p$ model.

\setcounter{figure}{2}
\begin{figure}[h]
\centering
\includegraphics[width=0.65\columnwidth]{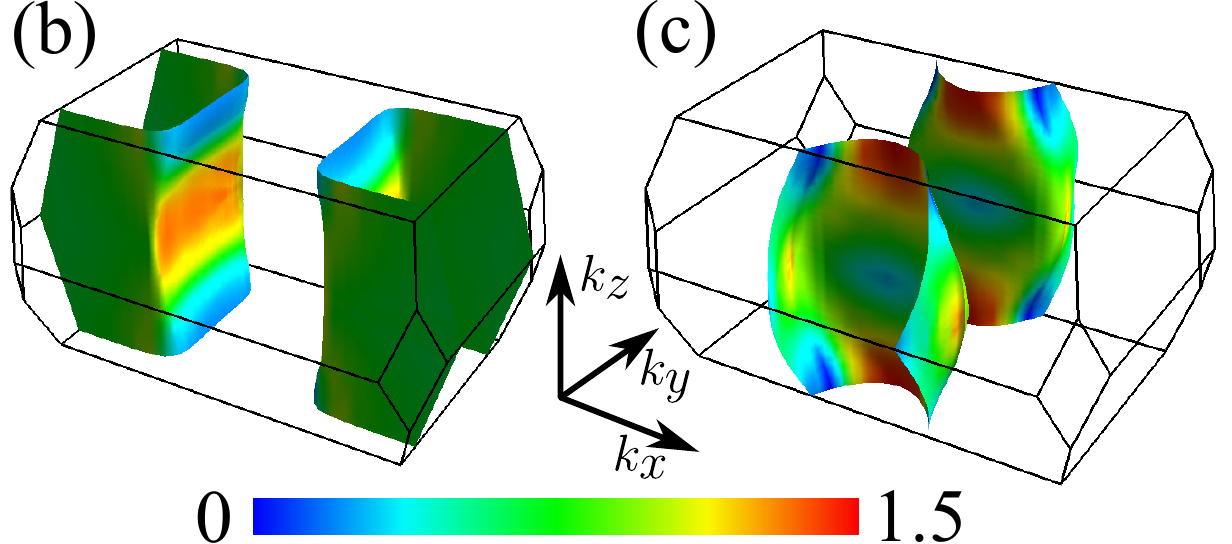}
\caption{
(b, c) Magnitude of the anomalous self-energy $\Delta_a(\bm{k}, i \pi T)$ on the $\alpha$ and $\beta$ Fermi surfaces, expressed in color.
Although point-node-like structures appeared at the corner of the $\alpha$-surface in the original text, correct point-node-like structures appear on the $\beta$-surface near the zone boundary of $k_z=2\pi/c$.
}
\label{e-fig3}
\end{figure}

\setcounter{figure}{5}
\begin{figure}[h]
\centering
\includegraphics[width=0.65\columnwidth]{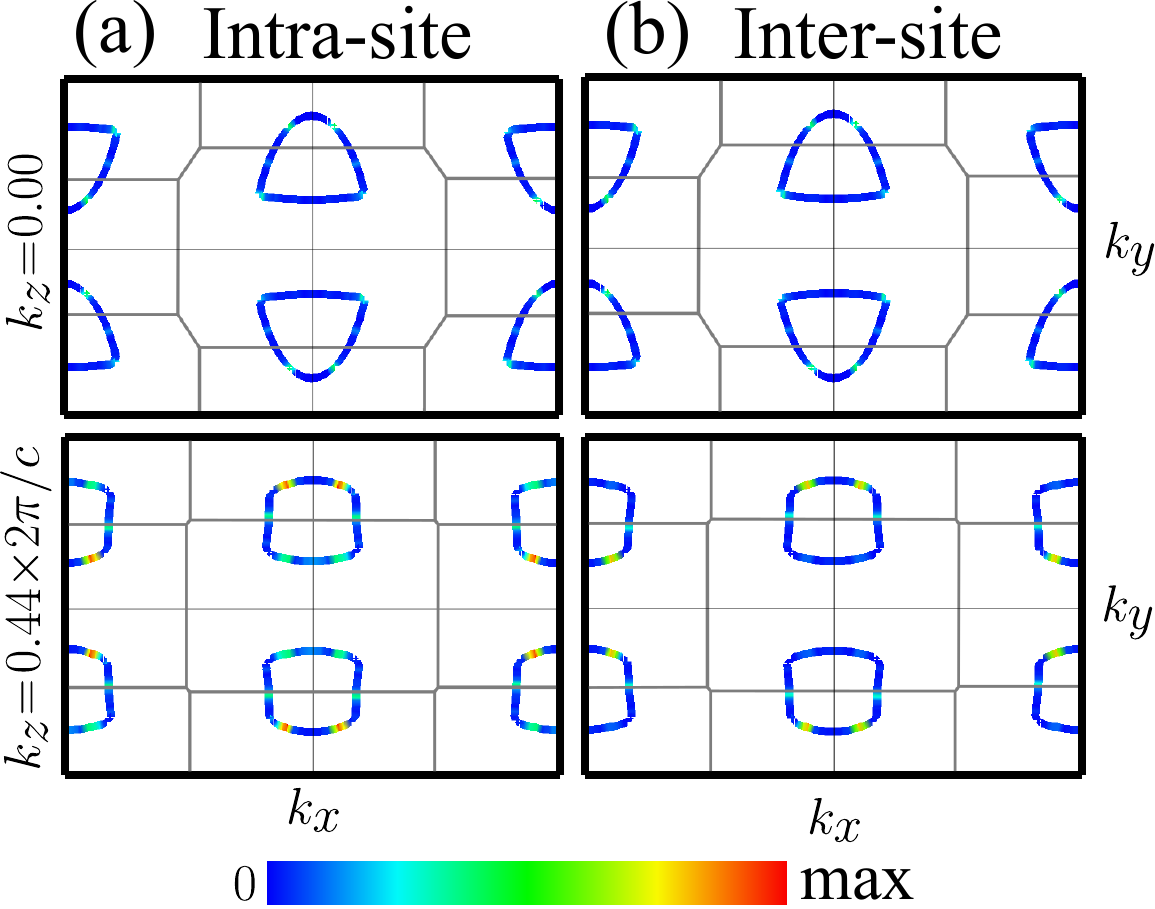}
\caption{
(a) Intra-site $\mathcal{F}_{ff}(\bm{k}, i \pi T) + \mathcal{F}_{f'f'}(\bm{k}, i \pi T)$ and (b) inter-site $\mathcal{F}_{ff'}(\bm{k}, i \pi T) + \mathcal{F}_{f'f}(\bm{k}, i \pi T)$ pairing amplitude on the $\beta$-surfaces in the $k_z = 0.0$ plane (upper panel) and $k_z = 0.44 \times 2 \pi / c$ plane (lower panel).
The gray lines indicate the Brillouin zone boundaries in each plane.
}
\label{e-fig6}
\end{figure}

\clearpage
\onecolumn

\makeatletter\def\fnum@figure{\figurename~S\thefigure}
\makeatletter\def\fnum@table{\tablename~S\thetable}

\centerline{{\large \textbf{Supplemental Material:}}}
\centerline{{\large \textbf{Unconventional \textit{s}-Wave Pairing with Point-Node--Like Gap Structure in UTe$_2$}}}

\vspace{10pt}

\centerline{{Shingo Haruna$^1$, Takuji Nomura$^{1,2}$, and Hirono Kaneyasu$^{1,3}$}}
\centerline{ \textit{
$^1$Department of Material Science, University of Hyogo, 3-2-1 Kouto, Ako, Hyogo 678-1297, Japan
}}
\centerline{ \textit{
$^2$Synchrotron Radiation Research Center, National Institutes for Quantum
}}
\centerline{ \textit{
Science and Technology, 1-1-1 Kouto, Sayo, Hyogo 679-5148, Japan
}}
\centerline{ \textit{
$^3$Center for Liberal Arts and Sciences, Sanyo-Onoda City University,
}}
\centerline{ \textit{
1-1-1 Daigakudori, Sanyo-Onoda, Yamaguchi 756-0884, Japan
}}

\vspace{10pt}

In this supplemental material, we provide the details of the band structure, our tight-binding model of $\mathrm{UTe_2}$, and the pairing interaction within the third-order perturbation theory.

\vspace{20pt}

\leftline{\textbf{I. Band Structure}}
We calculate the band structure of $\mathrm{UTe_2}$ within density functional theory (DFT) for the crystallographic parameters shown in Table S\ref{sTable1}, using the WIEN2k package \cite{sBlaha1}.
The crystal structure of this compound is an orthorhombic body-centered structure (Space group: $Immm$) with two uranium and four tellurium atoms in the primitive unit cell.
There are two crystallographically different Te sites in the unit cell, classified as Te1 and Te2 \cite{sHutanu1}.

\begin{figure}[h]
\centering
\includegraphics[width=0.8\columnwidth]{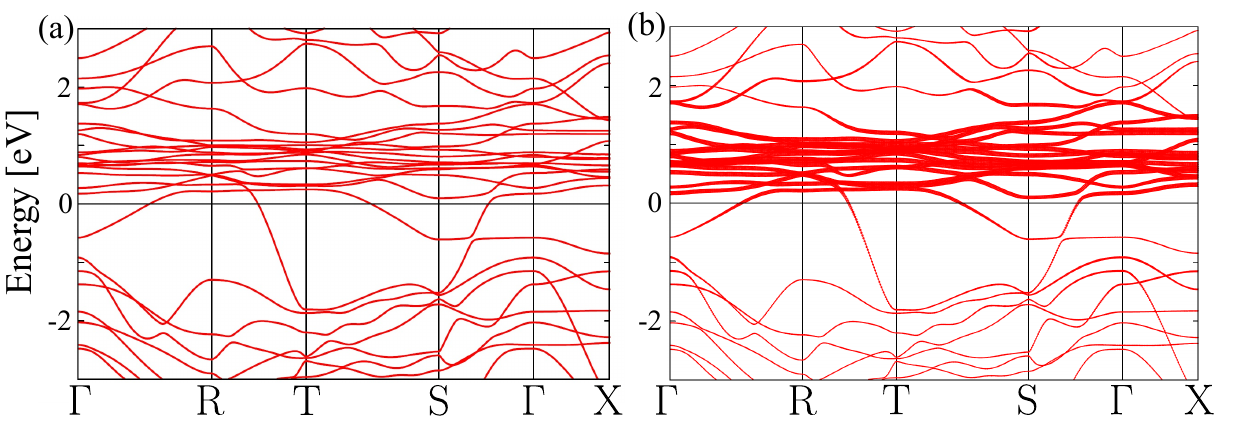}
\caption{Band structure produced by the GGA+U calculation using WIEN2k. (a) Energy dispersion on the symmetry lines.
(b) U-5$f$ electron weights on the bands are represented by line width.}
\label{sFig1-1}
\end{figure}

Our numerical calculation using the generalized gradient approximation with Hubbard $U$ (GGA+U) produces the band structure shown in Fig. S\ref{sFig1-1}, where we set $U=2.0$ eV.
In this calculation, we account for the spin--orbit coupling of uranium atoms and use $15\times15\times15$ $\bm{k}$-meshes and $R_\mathrm{MT} \times K_\mathrm{max} = 9.0$.
The GGA+U calculation reproduces two cylindrical Fermi surfaces and $f$-rich flat bands slightly above the Fermi level.
The calculated band structure agrees with previous theoretical studies \cite{sIshizuka2,sXu1}.

\begin{figure}[h]
\centering
\includegraphics[width=0.9\columnwidth]{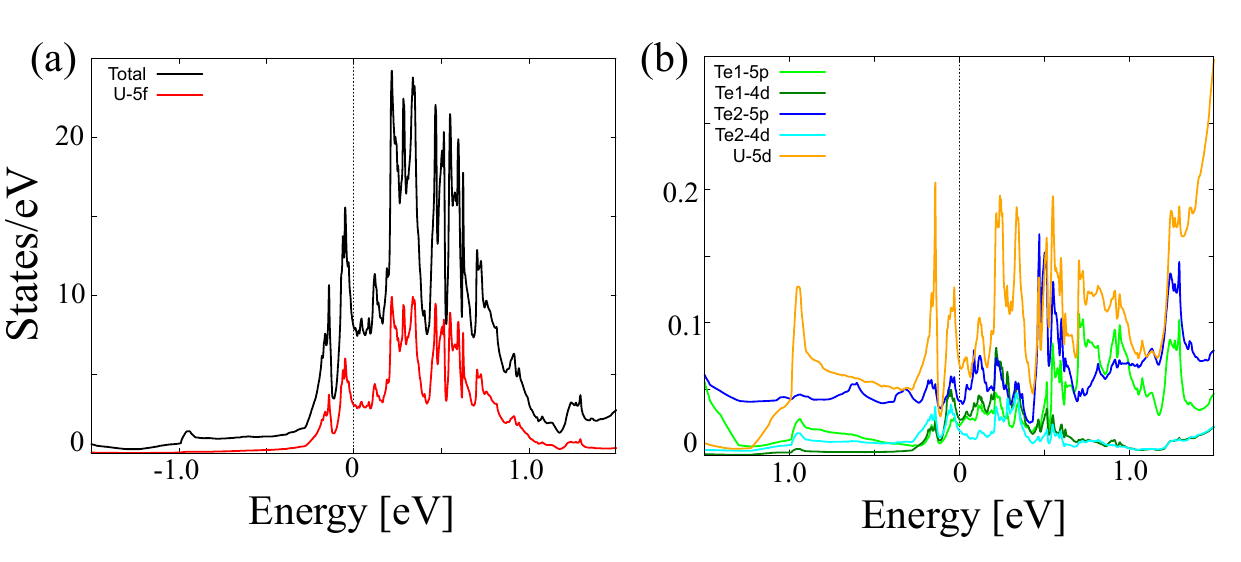}
\caption{Density of state (DOS) given by GGA+U calculation. 
(a) Total DOS and partial density of state (PDOS) of the U-5$f$ states.
(b) PDOS of U-6$d$ states and the Te-5$p$ and 4$d$ states.}
\label{sFig1-2}
\end{figure}

\begin{figure}[h]
\centering
\includegraphics[width=0.65\columnwidth]{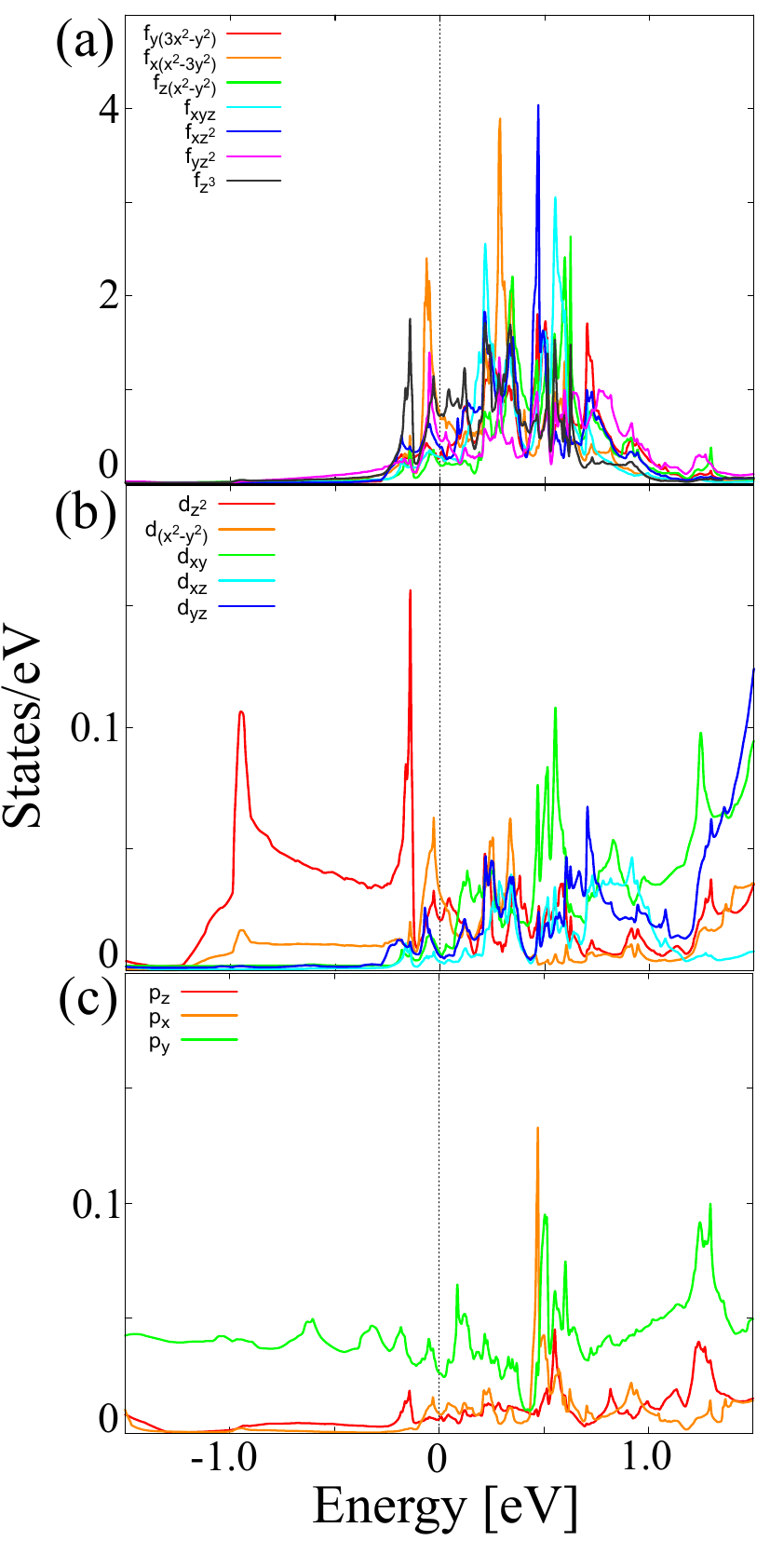}
\caption{(a), (b) and (c) show the PDOS for the U-5$f$, U-6$d$ and Te2-5$p$ states, respectively.}
\label{sFig1-3}
\end{figure}

Next, we calculate the density of states (DOS) and partial density of states (PDOS) for relevant orbitals, as shown in Figs. S\ref{sFig1-2} and S\ref{sFig1-3}.
The calculated DOS and PDOS for the U-5$f$, U-6$d$ and Te2-5$p$ orbital states are displayed in Fig. S\ref{sFig1-2}.
The 5$f$ state is dominant in the vicinity of the Fermi level (Fig. S\ref{sFig1-2} (a)), while the 6$d$ state of U and the 5$p$ state of Te2 take relatively large PDOS near the Fermi level (Fig. S\ref{sFig1-2} (b)).
To know the dominant orbital state near the Fermi level, we depict the PDOS for the U-5$f$, U-6$d$, and Te2-5$p$ states in Fig. S\ref{sFig1-3}.
Among the U-5$f$ states, the $f_{x(x^2-3y^2)}$ and $f_{z^3}$ states have the main contributions at the Fermi level. 
Conversely, among the U-6$d$ and Te2-5$p$ states, the dominant orbital states are $d_{x^2-y^2}$ and $p_y$, respectively.

\begin{table}[h]
\centering
\caption{Atomic positions of $\mathrm{UTe_2}$ in fractional coordinates. Lattice constants are $a = 4.123$ \AA, $b = 6.086$ \AA, and $c = 13.812$ \AA \cite{sHutanu1}.}
\label{sTable1}
\begin{tabular}{cccc}
\hline
Atom & $x$ & $y$ & $z$ \\ \hline
U & ~~~0.00000~~~ & ~~~0.00000~~~ & ~~~0.13473~~~ \\
Te1 & ~~~0.50000~~~ & ~~~0.00000~~~ & ~~~0.29799~~~ \\
Te2 & ~~~0.00000~~~ & ~~~0.25062~~~ & ~~~0.50000~~~ \\ \hline
\end{tabular}
\end{table}
\vspace{15pt}

\leftline{\textbf{II. $f$-$d$-$p$ Model}}
To construct the $f$-$d$-$p$ model for $\mathrm{UTe_2}$, we assume that the band structure consists of the three types of orbital states: U-5$f$, U-6$d$ and Te2-5$p$.
This assumption is the same as a previous study \cite{sIshizuka1}.
Furthermore, as the $f_{x(x^2-3y^2)}$ of U-5$f$, $d_{x^2-y^2}$ of U-6$d$ and $p_y$ of Te2-5$p$ orbital states have the dominant contributions to DOS, we use them to construct the effective tight-binding model.
Since there are two sets of U and Te2 atoms in the primitive unit cell, we have two sets of $f_{x(x^2-3y^2)}$, $d_{x^2-y^2}$, and $p_y$ orbitals.
To distinguish between the orbital states of the two sets of atoms, we use the notations $f, d, f', d', p, p'$ (see Table S\ref{sTable3})

\begin{table}[h]
\centering
\caption{Atom positions in fractional coordinates in the primitive unit cell and orbital states on each atom site.}
\label{sTable3}
\begin{tabular}{ccccc}
\hline
Atom & $x$ & $y$ & $z$ & Orbital State \\ \hline
U & ~~~$0.000$~~~ & ~~~$0.000$~~~ & ~~~$0.125$~~~ & $f, d$ \\
U & ~~~$0.000$~~~ & ~~~$0.000$~~~ & ~~~$0.875$~~~ & $f', d'$\\
Te2 & ~~~$0.500$~~~ & ~~~$0.250$~~~ & ~~~$0.000$~~~ & $p$ \\
Te2 & ~~~$0.000$~~~ & ~~~$0.250$~~~ & ~~~$0.500$~~~ & $p'$ \\ \hline
\end{tabular}
\end{table}

The $f$-$d$-$p$ model accounts for the U--U hoppings up to third-nearest neighbors and the U--Te2 and Te2--Te2 hoppings up to the nearest neighbors.
Including these five types of hoppings, the tight-binding Hamiltonian is given by
\begin{align}
\hat{H}_0 = \sum \limits_{\bm{k} \sigma} \hat{c}^\dagger_{\bm{k} \sigma} H_0(\bm{k}) \hat{c}_{\bm{k} \sigma}
\end{align}
where $\hat{c}_{\bm{k} \sigma} \equiv \begin{pmatrix}
c_{\bm{k}f \sigma} & c_{\bm{k}d \sigma} & c_{\bm{k}f' \sigma} & c_{\bm{k}d' \sigma} & c_{\bm{k}p \sigma} & c_{\bm{k}p' \sigma}
\end{pmatrix}^\mathrm{T}$.
$c_{\bm{k} l \sigma}$ and $c^\dagger_{\bm{k} l \sigma}$ represent the annihilation and creation operators for each orbital $l$ and spin $\sigma$, respectively.
The momentum dependency of the Hamiltonian $H_0(\bm{k})$ is formulated as
\begin{align}
H_0(\bm{k}) =
\begin{pmatrix}
\varepsilon_{f f}(\bm{k}) & \varepsilon_{f d}(\bm{k}) & \varepsilon_{f f'}(\bm{k}) & \varepsilon_{f d'}(\bm{k}) & \varepsilon_{f p}(\bm{k}) & \varepsilon_{f p'}(\bm{k}) \\
                                  & \varepsilon_{d d}(\bm{k}) & \varepsilon_{d f'}(\bm{k}) & \varepsilon_{d d'}(\bm{k}) & \varepsilon_{d p}(\bm{k}) & \varepsilon_{d p'}(\bm{k}) \\
                                  &                                   & \varepsilon_{f' f'}(\bm{k}) & \varepsilon_{f' d'}(\bm{k}) & \varepsilon_{f' p}(\bm{k}) & \varepsilon_{f' p'}(\bm{k}) \\
                                  &                                   &                                   & \varepsilon_{d' d'}(\bm{k}) & \varepsilon_{d' p}(\bm{k}) & \varepsilon_{d' p'}(\bm{k}) \\
                                  &              H.c.                &                                   &                                   & \varepsilon_{p p}(\bm{k}) & \varepsilon_{p p'}(\bm{k}) \\
                                  &                                   &                                   &                                   &                                   & \varepsilon_{p' p'}(\bm{k}) 
\end{pmatrix},
\end{align}
where $\varepsilon_{ij}(\bm{k})$ is the Hamiltonian element associated with the electron hopping from orbital $j$ to orbital $i$.
Matrix elements $\varepsilon_{ij}(\bm{k})$ for U--U hoppings are given by
\begin{align}
\varepsilon_{ff (dd)}(\bm{k}) &= \varepsilon_{f'f' (d'd')}(\bm{k}) = \varepsilon_0^{f(d)} + t_2^{f(d)} \Big( e^{-i k_x a} + e^{i k_x a} \Big), \\
\varepsilon_{fd}(\bm{k}) &= \varepsilon_{f'd'}(\bm{k}) = t_2^{fd} \Big( e^{-i k_x a} - e^{i k_x a} \Big), \\
\varepsilon_{ff'(dd')}(\bm{k}) &= t_1^{f(d)} e^{-i k_z c/4} + t_3^{f(d)} \Big[ e^{i(-k_x a/2 - k_y b/2)} + e^{i(-k_x a/2 + k_y b/2)} \notag \\
&~~~~~~~~~~~~~~~~~~~~~
+ e^{i( k_x a/2 - k_y b/2)}  + e^{i( k_x a/2 + k_y b/2)} \Big]e^{i k_z c/4}, \\
\varepsilon_{fd'}(\bm{k}) &= t_1^{fd} e^{-i k_z c/4} + t_3^{fd} \Big[ e^{i(-k_x a/2 - k_y b/2)}  + e^{i(-k_x a/2 + k_y b/2)} \notag \\
&~~~~~~~~~~~~~~~~~~~~~
- e^{i( k_x a/2 - k_y b/2)} - e^{i( k_x a/2 + k_y b/2)} \Big]e^{i k_z c/4}, \\
\varepsilon_{df'}(\bm{k}) &= t_1^{fd} e^{-i k_z c/4} - t_3^{fd} \Big[ e^{i(-k_x a/2 - k_y b/2)}  + e^{i(-k_x a/2 + k_y b/2)} \notag \\
&~~~~~~~~~~~~~~~~~~~~~
- e^{i( k_x a/2 - k_y b/2)} - e^{i( k_x a/2 + k_y b/2)} \Big]e^{i k_z c/4},
\end{align}
where $\varepsilon_0^{f(d)}$ is the one-particle energy of the $f \; (d)$ orbital.
Matrix elements for the hybridization between the U and Te2 sites are given by
\begin{align}
\varepsilon_{fp}(\bm{k}) &=
 t_4^{fp} \Big[ e^{i (- k_x a/2 + k_y b/4 - k_z c/8)} - e^{i (k_x a/2 + k_y b/4 - k_z c/8)} \Big], \\
\varepsilon_{dp}(\bm{k}) &=
 t_4^{dp} \Big[ e^{i (- k_x a/2 + k_y b/4 - k_z c/8)} + e^{i (k_x a/2 + k_y b/4 - k_z c/8)} \Big], \\
\varepsilon_{fp'}(\bm{k}) &=
- t_4^{fp} \Big[ e^{i (- k_x a/2 - k_y b/4 - k_z c/8)} - e^{i (k_x a/2 - k_y b/4 - k_z c/8)} \Big], \\
\varepsilon_{dp'}(\bm{k}) &=
- t_4^{dp} \Big[ e^{i (- k_x a/2 - k_y b/4 - k_z c/8)} + e^{i (k_x a/2 - k_y b/4 - k_z c/8)} \Big], \\
\varepsilon_{f'p}(\bm{k}) &=
 t_4^{fp} \Big[ e^{i (- k_x a/2 + k_y b/4 + k_z c/8)} - e^{i (k_x a/2 + k_y b/4 + k_z c/8)} \Big], \\
 \varepsilon_{d'p}(\bm{k}) &=
 t_4^{dp} \Big[ e^{i (- k_x a/2 + k_y b/4 + k_z c/8)} + e^{i (k_x a/2 + k_y b/4 + k_z c/8)} \Big], \\
\varepsilon_{f'p'}(\bm{k}) &=
- t_4^{fp} \Big[ e^{i (- k_x a/2 - k_y b/4 + k_z c/8)} - e^{i (k_x a/2 - k_y b/4 + k_z c/8)} \Big], \\
\varepsilon_{d'p'}(\bm{k}) &=
- t_4^{dp} \Big[ e^{i (- k_x a/2 - k_y b/4 + k_z c/8)} + e^{i (k_x a/2 - k_y b/4 + k_z c/8)} \Big].
\end{align}
Finally, $\varepsilon_{ij}(\bm{k})$ for Te2--Te2 hoppings are given by 
\begin{align}
\varepsilon_{pp}(\bm{k}) &= \varepsilon_{p'p'}(\bm{k}) = \varepsilon_{0}^p, \\
\varepsilon_{pp'}(\bm{k}) &= t_5^p \Big[ e^{-i k_y b/2} + e^{i k_y b/2} \Big].
\end{align}

\begin{figure}[h]
\centering
\includegraphics[width=0.8\columnwidth]{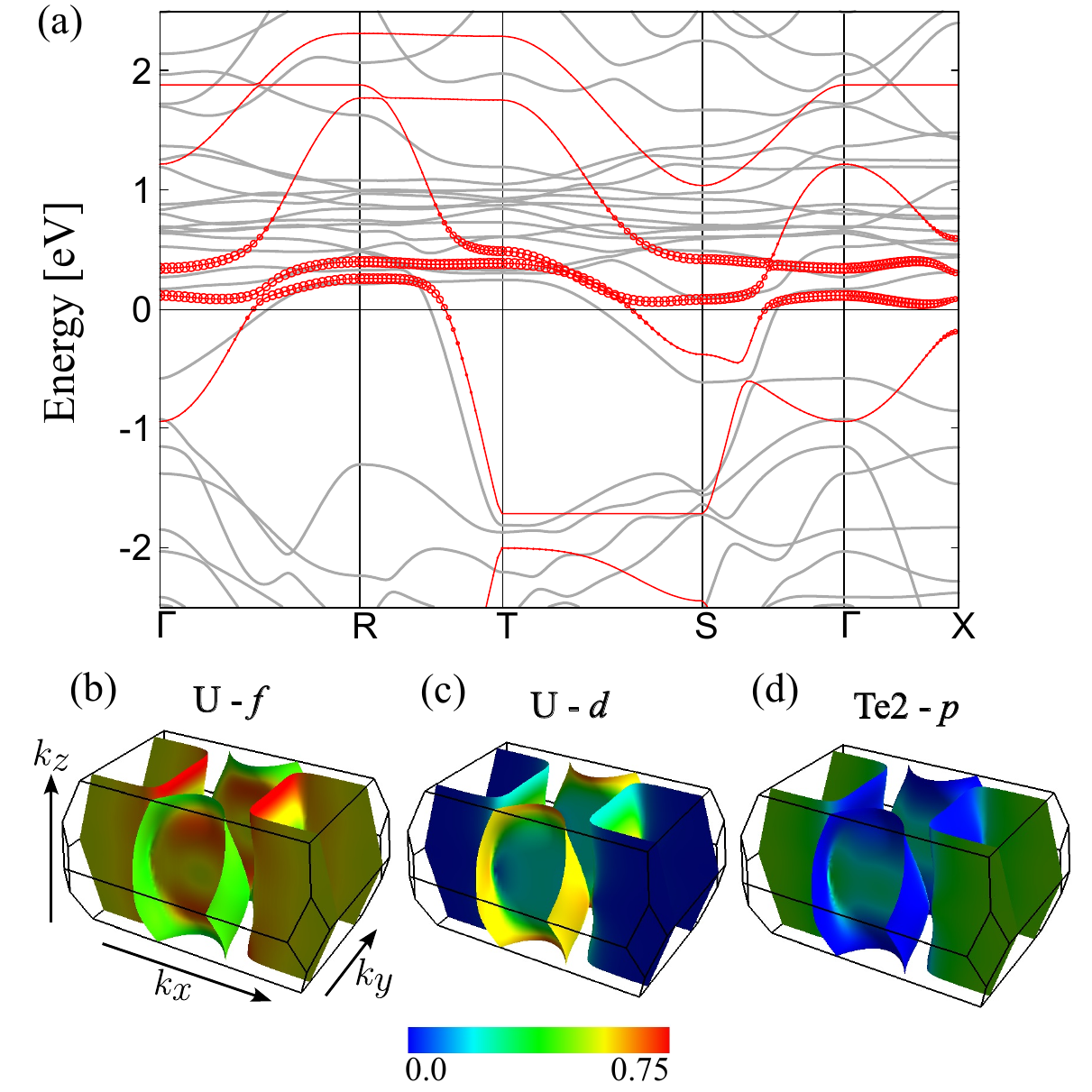}
\caption{(a) Band structure of the $f$-$d$-$p$ model.
The line width represents the weight of $f$ electrons and
the gray lines show the band structure given by WIEN2k.
(b) -- (d) Fermi surfaces are colored according to their weights of $f$, $d$ and $p$ electrons, respectively.}
\label{sFig2-2}
\end{figure}

In this study, the one-particle energy and the hopping parameters are set as in Table S\ref{sTable2}.
By the diagonalization of the tight-binding Hamiltonian with these hopping parameters, the band structure and the quasi-two-dimensional Fermi surfaces are reproduced as depicted in Fig. S\ref{sFig2-2}.
The $f$-$d$-$p$ model reproduces the $f$-rich flat bands slightly above the Fermi level, consistent with the WIEN2k calculation. 
The computed Fermi surfaces agree with the de Haas--van Alphen measurements \cite{sAoki1} and angle-resolved photoemission spectroscopy experiments \cite{sMiao1,sFujimori1}.
Fermi surface portions nearly parallel to the $k_x$ axis mainly consist of $f$ and $p$ components, whereas Fermi surface portions nearly parallel to the $k_y$ axis mainly consists of the $f$ and $d$ components.

\begin{table}[h]
\centering
\caption{The hopping parameters for the $f$-$d$-$p$ model in units of eV.
$\varepsilon^f_0, \varepsilon^d_0$ and $\varepsilon^p_0$ are set to $\varepsilon^f_0 = 0.2,\; \varepsilon^d_0 = 0.9,\; \varepsilon^p_0 = -1.8$.
$t_1, t_2$, and $t_3$ represent the U--U hoppings for the first, second and third nearest neighbors, respectively.
$t_4$ and $t_5$ denote the U--Te2 and Te2--Te2 hoppings for the first nearests, respecively.}
\label{sTable2}
\begin{tabular}{cccccccc}
\hline
~~~~~~~~~~ &    Atom    &~~~$t^f$~~~&~~~$t^d$~~~&~~~$t^{fd}$~~~&~~~$t^{fp}$~~~&~~~$t^{dp}$~~~&~~~$t^p$~~~\\
\hline
$t_1$   & U--U       & $-0.1$       & $0.275$        & $0.25$        & ---       & ---           & ---        \\
$t_2$   & U--U       & $-0.025$    & $-0.5$         & $0.1$          & ---       & ---           & ---        \\
$t_3$   & U--U       & $0.001$     & $0.15$         & $0.05$         & ---       & ---           & ---        \\
$t_4$   & U--Te2    & ---           & ---            & ---             & $-0.2$    & $-0.35$     & ---        \\
$t_5$   & Te2--Te2 & ---           & ---            & ---             & ---        & ---          & $1.8$      \\
\hline
\end{tabular}
\end{table}

\vspace{15pt}

\leftline{\textbf{III. Pairing Interaction}}
In our formulation, the pairing interaction $V^{S/T}_{aa'}$ in the band representation is related to the pairing interaction $V^{S/T}_{l_1 l_2, l_3 l_4}$ in the orbital representation as follows \cite{sYanase1,sNomura1}:
\begin{align}
V^{S/T}_{aa'}(k, k') = \sum_{l_1 l_2 l_3 l_4} U^*_{l_1 a}(\bm{k}) U^*_{l_2 a}(-\bm{k}) V^{S/T}_{l_1 l_2, l_3 l_4}(k, k') U_{l_3 a'}(-\bm{k}') U_{l_4 a'}(\bm{k}'),
\end{align}
where $V^{S/T}_{aa'}(k, k')$ is the pairing interaction for spin-singlet/triplet channels, and $l_i$ and $a$ are orbital and band indices, respectively.
$U_{l a}(\bm{k})$ denotes the elements of the diagonalization matrix of the tight-binding Hamiltonian, that is,
\begin{align}
c_{\bm{k} l \sigma} = \sum \limits_a U_{l a}(\bm{k}) c_{\bm{k} a \sigma}.
\end{align}
$c_{\bm{k} l \sigma}$ and $c_{\bm{k} a \sigma}$ are the annihilation operators in the orbital and band representation, respectively.
The pairing interaction $V^{S}_{l_1 l_2, l_3 l_4}(k, k')$ and $V^{T}_{l_1 l_2, l_3 l_4}(k, k')$ are given by
\begin{align}
V^{S}_{l_1 l_2, l_3 l_4}(k, k') &= \frac{1}{2} \sum \limits_{\sigma_1 \sigma_2 \sigma_3 \sigma_4} V_{\zeta_1 \zeta_2, \zeta_3 \zeta_4}(k, k')
\left( i \sigma^y \right)^\dagger_{\sigma_1 \sigma_2} \left( i \sigma^y \right)_{\sigma_3 \sigma_4}, \\
V^{T}_{l_1 l_2, l_3 l_4}(k, k') \delta_{\mu \nu}&= \frac{1}{2} \sum \limits_{\sigma_1 \sigma_2 \sigma_3 \sigma_4} V_{\zeta_1 \zeta_2, \zeta_3 \zeta_4}(k, k')
\left( i \sigma^\mu \sigma^y \right)^\dagger_{\sigma_1 \sigma_2} \left( i \sigma^\nu \sigma^y \right)_{\sigma_3 \sigma_4},
\end{align}
where $\zeta_i$ denotes the orbital-spin combined index defined as $\zeta_i \equiv (l_i, \sigma_i)$.

The effective pairing interaction is perturbatively expanded to third order as follows:
\begin{align}
V_{\zeta_1 \zeta_2, \zeta_3 \zeta_4}(k, k') &= V_{\zeta_1 \zeta_2, \zeta_3 \zeta_4}^{(a)}(k, k')
+ V_{\zeta_1 \zeta_2, \zeta_3 \zeta_4}^{(b)}(k, k')
+ V_{\zeta_1 \zeta_2, \zeta_3 \zeta_4}^{(c)}(k, k')
+ V_{\zeta_1 \zeta_2, \zeta_3 \zeta_4}^{(d)}(k, k') \notag \\
&~~~~~~~~~
+ V_{\zeta_1 \zeta_2, \zeta_3 \zeta_4}^{(e)}(k, k')
+ V_{\zeta_1 \zeta_2, \zeta_3 \zeta_4}^{(f)}(k, k')
+ V_{\zeta_1 \zeta_2, \zeta_3 \zeta_4}^{(g)}(k, k').
\label{int}
\end{align}
The diagrammatic representations of $V_{\zeta_1 \zeta_2, \zeta_3 \zeta_4}^{(a)}(k, k'),\; \ldots, \; V_{\zeta_1 \zeta_2, \zeta_3 \zeta_4}^{(g)}(k, k')$ are shown in Fig. 2 of the main text.
Each term of Eq. (\ref{int}) is analytically given by
\begin{align}
V_{\zeta_1 \zeta_2, \zeta_3 \zeta_4}^{(a)}(k, k') &=
\Gamma^{(0)}_{\zeta_1 \zeta_2, \zeta_3 \zeta_4}, \label{double1} \\
V_{\zeta_1 \zeta_2, \zeta_3 \zeta_4}^{(b)}(k, k') &=
- \sum \limits_{ \{ \xi_i \} } \chi^{(0)}_{\xi_2 \xi_4, \xi_1 \xi_3}(k - k') \Gamma^{(0)}_{\zeta_1 \xi_2, \xi_3 \zeta_4} \Gamma^{(0)}_{\xi_1 \zeta_2, \zeta_3 \xi_4}, \\
V_{\zeta_1 \zeta_2, \zeta_3 \zeta_4}^{(c)}(k, k') &=
\sum \limits_{ \{ \xi_i \} } \sum \limits_{ \{ \eta_i \} }
\chi^{(0)}_{\xi_2 \xi_4, \xi_1 \xi_3} (k - k') \chi^{(0)}_{\eta_2 \eta_4, \eta_1 \eta_3} (k - k') \notag \\
&~~~\times
\Gamma^{(0)}_{\zeta_1 \xi_2, \xi_3 \zeta_4} \Gamma^{(0)}_{\xi_1 \eta_2, \eta_3 \xi_4} \Gamma^{(0)}_{\eta_1 \zeta_2, \zeta_3 \eta_4}, \\
V_{\zeta_1 \zeta_2, \zeta_3 \zeta_4}^{(d)}(k, k') &=
\frac{T}{N} \sum \limits_{k_1} \sum \limits_{ \{ \xi_i \} } \sum \limits_{ \{ \eta_i \} }
\chi^{(0)}_{\xi_1 \xi_3, \xi_4 \xi_2} (- k + k_1) \mathcal{G}^{(0)}_{\eta_2 \eta_1}(k_1) \notag \\
&~~~\times
\mathcal{G}^{(0)}_{\eta_4 \eta_3}(k_1 - k + k') 
\Gamma^{(0)}_{\xi_1 \eta_3, \xi_2 \zeta_4} \Gamma^{(0)}_{\zeta_1 \xi_3, \eta_2 \xi_4} \Gamma^{(0)}_{\eta_1 \zeta_2, \zeta_3 \eta_4}, \\
V_{\zeta_1 \zeta_2, \zeta_3 \zeta_4}^{(e)}(k, k') &=
\frac{T}{N} \sum \limits_{k_1} \sum \limits_{ \{ \xi_i \} } \sum \limits_{ \{ \eta_i \} }
\chi^{(0)}_{\xi_2 \xi_4, \xi_3 \xi_1} (k + k_1) \mathcal{G}^{(0)}_{\eta_1 \eta_2}(k_1) \notag \\
&~~~\times
\mathcal{G}^{(0)}_{\eta_3 \eta_4}(k_1 + k - k') 
\Gamma^{(0)}_{\zeta_1 \eta_2, \eta_3 \zeta_4} \Gamma^{(0)}_{\eta_4 \xi_2, \zeta_3 \xi_1} \Gamma^{(0)}_{\xi_4 \zeta_2, \xi_3 \eta_1}, \\
V_{\zeta_1 \zeta_2, \zeta_3 \zeta_4}^{(f)}(k, k') &=
\frac{1}{2} \frac{T}{N} \sum \limits_{k_1} \sum \limits_{ \{ \xi_i \} } \sum \limits_{ \{ \eta_i \} }
\phi^{(0)}_{\xi_1 \xi_2, \xi_4 \xi_3} (k + k_1) \mathcal{G}^{(0)}_{\eta_1 \eta_2}(k_1) \notag \\
&~~~\times
\mathcal{G}^{(0)}_{\eta_3 \eta_4}(k_1 + k - k') 
\Gamma^{(0)}_{\xi_1 \xi_2, \eta_3 \zeta_4} \Gamma^{(0)}_{\zeta_1 \eta_2, \xi_3 \xi_4} \Gamma^{(0)}_{\eta_4 \zeta_2, \zeta_3 \eta_1}, \label{double2} \\
V_{\zeta_1 \zeta_2, \zeta_3 \zeta_4}^{(g)}(k, k') &=
\frac{1}{2} \frac{T}{N} \sum \limits_{k_1} \sum \limits_{ \{ \xi_i \} } \sum \limits_{ \{ \eta_i \} }
\phi^{(0)}_{\xi_2 \xi_1, \xi_3 \xi_4} (- k + k_1) \mathcal{G}^{(0)}_{\eta_2 \eta_1}(k_1) \notag \\
&~~~\times
\mathcal{G}^{(0)}_{\eta_4 \eta_3}(k_1 - k + k') 
\Gamma^{(0)}_{\zeta_1 \eta_3, \eta_2 \zeta_4} \Gamma^{(0)}_{\xi_1 \xi_2, \zeta_3 \eta_4} \Gamma^{(0)}_{\eta_1 \zeta_2, \xi_3 \xi_1}, \label{double3} 
\end{align}
with
\begin{align}
\chi^{(0)}_{\zeta_1 \zeta_2, \zeta_3 \zeta_4}(Q) &\equiv - \frac{T}{N} \sum \limits_{k} \mathcal{G}^{(0)}_{\zeta_3 \zeta_1}(k) \mathcal{G}^{(0)}_{\zeta_4 \zeta_2}(k + Q), \\
\phi^{(0)}_{\zeta_1 \zeta_2, \zeta_3 \zeta_4}(Q) &\equiv - \frac{T}{N} \sum \limits_{k} \mathcal{G}^{(0)}_{\zeta_3 \zeta_1}(k) \mathcal{G}^{(0)}_{\zeta_4 \zeta_2}(Q - k),
\end{align}
where $N$ is the number of $\bm{k}$-mesh points and the summation with $\{ \xi_i \}$ or $\{ \eta_i \}$ denotes that with respect to the orbital and spin states.
The argument $Q$ represents the four-dimensional momentum $(\bm{Q}, i \Omega_n)$, where $\Omega_n \equiv 2n \pi T$ is the bosonic Matsubara frequencies.
The factor of $1/2$ in Eq. (\ref{double2}) and (\ref{double3}) is necessary to avoid double counting of equivalent diagrams.
When these expressions are used in the Eliashberg equation Eq. (3) of the main text, $V^{(a)}_{\zeta_1 \zeta_2, \zeta_3 \zeta_4}(k, k')$ has to be multiplied by a factor of $1/2$ to avoid double counting of the first order $V^{(a)}_{\zeta_1 \zeta_2, \zeta_3 \zeta_4}(k, k')$.
$\Gamma^{(0)}_{\zeta_1 \zeta_2, \zeta_3 \zeta_4}$ is the antisymmetrized bare vertex given by
\begin{align}
\Gamma^{(0)}_{\zeta_1 \zeta_2, \zeta_3 \zeta_4} = I_{\zeta_1 \zeta_2, \zeta_3 \zeta_4} - I_{\zeta_1 \zeta_2, \zeta_4 \zeta_3},
\end{align}
where $I_{\zeta_1 \zeta_2, \zeta_3 \zeta_4}$ is represented as follows:
\begin{align}
I_{(l_1 \sigma_1)(l_2 \sigma_2), (l_3 \sigma_3)(l_4 \sigma_4)} =
\begin{cases}
U_l & (l_1=l_2=l_3=l_4=l, \; \sigma_1 = \sigma_4 \neq \sigma_2 = \sigma_3) \\
0 & (\mathrm{otherwise})
\end{cases}.
\end{align}
In the $f$-$d$-$p$ model, as the on-site Coulomb repulsion is set only at the $f$ orbital, $U_l$ is given by
\begin{align}
U_l =
\begin{cases}
U & (\mathrm{for}\; l = f, f') \\
0 & (\mathrm{otherwise})
\end{cases},
\end{align}
where the parameter $U$ is set to 1.75 eV or 1.50 eV in this study.

%

\clearpage


\end{document}